\documentclass[aps,10pt,prd,notitlepage]{revtex4-1}
\usepackage[utf8]{inputenc}
\usepackage{amsmath,amssymb,bm,epsfig}
\usepackage{color}
\usepackage{hyperref} 
\usepackage{natbib}
\usepackage{ulem}
\usepackage{graphicx}
\usepackage{xcolor}
\usepackage{pifont}
\usepackage{amsfonts}
\usepackage{mathrsfs}
\usepackage{cancel}
\usepackage{bm}
\usepackage{bigints}
\usepackage{float}
\usepackage{subfloat}
\usepackage{caption}
\usepackage{subcaption}

\def\be {\begin{equation}}
\def\ee {\end{equation}}
\def\bea {\begin{eqnarray}}
\def\eea {\end{eqnarray}}
\def\bc {\begin{center}}
\def\ec {\end{center}}
\def\nn {\nonumber}

\def\mn {\mu\nu}
\def\vp{\varpi}
\def\({\left(}
\def\){\right)}
\def\[{\left[}
\def\]{\right]}

\DeclareGraphicsExtensions{.jpg,.pdf,.eps}

\makeatletter
\def\thickhline{%
  \noalign{\ifnum0=`}\fi\hrule \@height \thickarrayrulewidth \futurelet
   \reserved@a\@xthickhline}
\def\@xthickhline{\ifx\reserved@a\thickhline
               \vskip\doublerulesep
               \vskip-\thickarrayrulewidth
             \fi
      \ifnum0=`{\fi}}
\makeatother

\newlength{\thickarrayrulewidth}
\usepackage{makecell}
\usepackage{boldline}

\begin{document}
\title{Covariant formulation of gluon self-energy in presence of  ellipsoidal  anisotropy}

\author{Ritesh Ghosh$^{a,b}$}
\email{ritesh.ghosh@saha.ac.in}
\author{Bithika Karmakar$^{a,b}$}
\email{bithika.karmakar@saha.ac.in}
\author{Arghya Mukherjee$^{a,c}$}
\email{arbp.phy@gmail.com}
\affiliation{$^a$Saha Institute of Nuclear Physics, 1/AF Bidhannagar, Kolkata - 700064, India}
\affiliation{$^b$Homi Bhabha National Institute, 2nd floor, BARC Training School Complex, Anushaktinagar, Mumbai, Maharashtra 400094, India}
%\author{Mahatsab Mandal$^e$}
%\email{mahatsab@gmail.com}
%\author{ Pradip Roy$^{b,d}$}
%\email{pradipk.roy@saha.ac.in}
\affiliation{$^c$School of Physical Sciences, National Institute of Science Education and Research, HBNI, Jatni 752050, India}

%+++++++++++++++++++++++++++++++++++++++++++++++++++++++++++++++++++++++++++++++++++++++++++++++++++++++++++++++++++++
\begin{abstract}

In this work, a covariant formulation of the  gluon self-energy in presence of ellipsoidal anisotropy  is considered. It is shown that the general structure of the gluon self-energy can be written in terms of six linearly independent projection tensors. Similar to the spheroidal anisotropy,   mass scales can be introduced for  each of the collective modes considering the static limits. 
With a simplified  ellipsoidal generalization of the  Romatschke-Strickland  form, the angular dependencies of the mass scales are studied. It is observed that, compared to the spheroidal case,  additional unstable mode  may appear in presence of  ellipsoidal anisotropy depending upon the choice of the parameters.

\end{abstract}

\maketitle
%+++++++++++++++++++++++++++++++++++++++++++++++++++++++++++++++++++++++++++++++++++++++++++++++++++++++++++++++++++
%
\section{Introduction}\label{sec.intro}

Over the last few decades, 
the plethora of knowledge gained through the  nonperturbative lattice QCD simulations \cite{Ding:2015ona,Ratti:2018ksb,Bazavov:2019lgz}, perturbative hard thermal loop calculations \cite{Braaten:1989mz,Andersen:2004fp,Haque:2014rua,Su:2015esa}, effective hydrodynamical modeling \cite{Gale:2013da,Jeon:2015dfa,Jaiswal:2016hex} as well as AdS/CFT inspired studies \cite{Son:2007vk,Nastase:2007kj,Peschanski:2008xn,Sadeghi:2013zma} have paved the way of achieving a remarkable advancement in our understanding of the experimental data available from the Relativistic Heavy Ion Collider (RHIC) facility at Brookhaven National Lab and the  Large Hadron Collider  (LHC) facility at European Center for Nuclear Research (CERN)\cite{Stock:2010hoa,Busza:2018rrf}. The concerted efforts from different heavy ion  research  communities suggest that the deconfined quark-gluon plasma(QGP) matter produced in the ultrarelativistic  heavy ion collision experiments  is most likely to possess substantial  deviation  from  perfect local isotropic equilibrium \cite{Strickland:2013uga}. In fact, soon after the initial nuclear impact, very large pressure anisotropy is expected in the center of the fireball with even larger  anisotropy prevailing in the cooler regions of the plasma \cite{Strickland:2014pga}. A promising way for hydrodynamical modeling  of such highly momentum space anisotropic system  is to consider the framework of anisotropic hydrodynamics \cite{Alqahtani:2017jwl,Alqahtani:2017mhy}. On the other hand, incorporating  certain classes of  anisotropic momentum distributions in the framework of hard thermal loop perturbation theory,  the nonequilibrium plasma properties  can  be extracted by studying  the collective modes of the  quasipartons   \cite{Mrowczynski:2004kv,Mrowczynski:2000ed}. A suitable parametrization for such   distributions  can be achieved following Refs.~\cite{Romatschke:2003ms,Romatschke:2004jh,Schenke:2006fz} where the one particle isotropic momentum space  distribution function  is deformed  by introducing a  directional dependency. This particular Romatschke-Strickland (RS) form has  been 
widely used in different phenomenological applications such as photon and dilepton production from anisotropic QGP \cite{Bhattacharya:2008up,Bhattacharya:2008mv,Bhattacharya:2015ada}, anisotropic heavy quark potential \cite{Nopoush:2017zbu}, bottomonia suppression \cite{Krouppa:2016jcl}, plasma wakes \cite{Mandal:2012wi,Mandal:2013jla}, nuclear modification factor \cite{Mandal:2011xn}, quasiparticle descriptions  of particle production \cite{Chandra:2016dwy} and so on. 

An important aspect of considering the nonequilibrium momentum distributions in QGP medium is the occurrence of kinetic instabilities. These are, in simple terms, the collective modes that possess  a positive imaginary part in their mode  frequencies resulting an exponential growth in the chromomagneic and chromoelectric fields.  The existence of such  Chromo-Weibel instabilities \cite{Strickland:2006cv} can influence the thermalization and  isotropization  of the medium \cite{Rebhan:2004ur}. A recent review with a pedagogical introduction to the required theoretical tools for such studies can be found in Ref. \cite{Mrowczynski:2016etf}. In general, how many  unstable modes are possible depends on the choice of the parameters. For example, in case of spheroidal anisotropy with Romatschke-Strickland parametrized form, the number of the possible unstable modes differs for the prolate and the oblate case which are obtained considering the negative and positive values for the anisotropy parameter respectively \cite{Romatschke:2003ms}. Moreover, there will be a directional dependence as well. In other words, depending on the angle of propagation with respect to the anisotropy direction, a stable collective mode may become unstable and can give rise to additional instabilities. 

As already mentioned, the large momentum space anisotropy in early stages can be efficiently incorporated in the aHydro framework. This is because, unlike usual viscous hydrodynamic set up, here, the dominant anisotropic contributions to the distribution functions are captured  in the leading order of hydrodynamic expansion.  The second order anisotropic hydrodynamics as developed in \cite{Bazow:2013ifa} can take into account arbitrary transverse expansion and is consistent with the traditional dissipative hydrodynamics approach in the limiting case of small anisotropy. However, as argued in Ref. \cite{Tinti:2013vba},  the azimuthally symmetric ansatz in such approaches involving a single anisotropy parameter can further be generalized in a systematic way. More specifically, in  the original anisotropic hydrodynamic setup, the leading order local rest frame distribution is taken to be of RS form and thus the two components of the  pressure in the transverse plane can  be different only after including the second order corrections. A generalization to include three different pressure components in the leading order of hydrodynamic expansion has been developed in Refs. \cite{Tinti:2013vba,Nopoush:2014pfa,Alqahtani:2015qja} and recently  has further been generalized in Ref. \cite{Nopoush:2019vqc}. In the hard loop approach too, the generalized ellipsoidal distribution has  been considered in the parton self-energy studies \cite{Kasmaei:2016apv} and implemented in phenomenological applications of photon and dilepton production \cite{Kasmaei:2019ofu,Kasmaei:2018oag}.  
  Obviously, the consideration of an ellipsoidal anisotropy  increases  the number of  parameters compared to the spheroidal case, as, apart from one additional anisotropy parameter dependence,  the collective properties of the partons in this case also possess azimuthal angular dependency.
Recently,  the  gluonic unstable mode with an ellipsoidal anisotropic momentum distribution has been reported for the first time in Ref. \cite{Kasmaei:2018yrr}. 
It is observed  that  the  growth rate of the unstable modes can become several times larger than the spheroidal case for certain propagation directions. Also the number of possible unstable modes are direction dependent. 
A convenient method to describe such directional dependency is to consider the static limits of the collective modes. In that case, one introduces mass scales corresponding to each of the collective modes and studies their angular dependency. The occurrence of a negative value in the  mass scale in fact signifies the presence of an instability in the corresponding mode which are well studied in the spheroidal case. However, for the ellipsoidal case, such mass scales could not be defined  in the formalism adopted in  \cite{Kasmaei:2018yrr} where the general structure of the gluon polarization has not been considered.  An important application of gluon self energy lies in the determination of the perturbative part of the  heavy quark potential \cite{Dumitru:2007hy,Dumitru:2009fy,Nopoush:2017zbu} which requires a covariant formulation of the general structure. 

The  primary objective of this work is  to formulate a  general structure for the gluon polarization in presence of ellipsoidal momentum space anisotropy in a covariant way and  study the occurrence of the unstable modes in a  similar fashion as done in case of spheroidal momentum space anisotropy.   A suitable approach for constructing the tensor basis for gluon self energy  is to choose the maximum possible mutually orthogonal set. The  choice becomes useful for the derivation of the effective propagator where  contractions among the basis tensors are involved. Once  the effective gluon propagator is derived, one can obtain the  collective modes  from its  pole.   
It should be mentioned here that the gluon collective modes can also be obtained by solving the characteristic equation without requiring any consideration  of the general structure. In fact, this is the procedure  adopted in Ref.\cite{Kasmaei:2018yrr}.   However, the advantage of considering the general structure is  that, the collective modes are expressed in terms of coordinate independent form factors which is essential for introducing  the mass scales. In fact, the nontrivial  angular dependence  in the hard-thermal loop resummed potential  enters  through these mass scales \cite{Nopoush:2017zbu}.    

The article is organized as follows. We first describe the formalism for constructing the general structure for the gluon polarization function in presence of the ellipsoidal anisotropy. The derived structure  is subsequently used to obtain the effective propagator and the collective modes of the gluon. Considering to a simple  ellipsoidal generalization of the RS form, the numerical results for the form factors and the mass scales are presented in the following section. Finally we conclude with a brief summary   of the presented  work. The following convention is used throughout the article: metric $\eta^{\mn}=\mbox{diag}(1,-1,-1,-1)$ and a general momentum four vector $Q^\mu=(q^0,\bm q)$ with $Q^2$ representing  $\eta_{\mn} Q^\mu Q^\nu=Q\cdot Q$ and $q=|{\bm q}|$.

\section{Formalism}\label{sec.formalism}

We consider a thermal QCD medium with a general  anisotropic momentum  distribution characterized by two independent  four vectors  $a^\mu$ and $b^\mu$ having unit norms. These two,  together with the normalized heat bath velocity ($u^\mu$) and  the gluon  momentum  ($P^\mu$),  can be used to form  a set of ten  independent symmetric  tensors so that  the symmetric gluon polarization tensor can be expressed as a linear combination of them. A simple choice for this purpose may be the set of tensors  $P^\mu P^\nu$, $u^\mu u^\nu$, $b^\mu b^\nu$, $a^\mu a^\nu$, $P^\mu u^\nu+P^\nu u^\mu$, $P^\mu b^\nu+P^\nu b^\mu$, $P^\mu a^\nu+P^\nu a^\mu$, $u^\mu b^\nu+b^\mu u^\nu$, $u^\mu a^\nu+u^\nu a^\mu$ and $b^\mu a^\nu+a^\mu b^\nu$. Notice that, along with the four vectors, we have not considered the metric $\eta^{\mn}$ as usually done in isotropic  or even in  anisotropic case with one anisotropy direction. This is because, in case of ellipsoidal anisotropy, the metric  itself no longer remains an independent tensor and can be expressed as a linear combination of the chosen set. Now, the constraints from the transversality condition $P^\mu \Pi_{\mn}=0 $ further reduce the number of independent basis tensors to six. In the  rest frame of the heat bath with $u^\mu=(1,0,0,0)$,  one of the anisotropy directions  can be taken along  z,  say  $b^\mu=(0,0,0,1)$, whereas the other  anisotropy direction can be assumed to lie in the $xz$ plane without any loss of generality. In the following we discuss a convenient method to obtain the basis tensors in a systematic way. 

Let us first consider the general structure of the  gauge boson self-energy in vacuum that  can be written as
\bea
\Pi^{\mn}=\left(\eta^{\mn}-\frac{P^\mu P^\nu}{P^2}\right) \Pi(P^2)= V^{\mn} \Pi(P^2).
\eea
Using the tensor $V^{\mn}$,  we obtain
$\tilde u^\mu=V^{\mn}u_\nu$ which can be used to construct the first basis  tensor given by 
\bea
A^{\mn}=\frac{\tilde u^\mu \tilde u^\nu}{\tilde u^2}.
\eea
Now, it is useful to define the  $A^{\mn}$ subtracted part of $V^{\mn}$ as $U^{\mn}=V^{\mn}-A^{\mn}$ which can be used to  obtain $\tilde b^\mu$ defined as $\tilde b^\mu=U^{\mn}b_\nu$ such that it  becomes orthogonal to $\tilde u^\mu$ by construction. Similar to the earlier  case, we obtain our  second  basis tensor using $\tilde b^\mu$ as
\bea
B^{\mn}=\frac{\tilde b^\mu \tilde b^\nu}{\tilde b^2}.
\eea
Another symmetric tensor that can be constructed intuitively using  $\tilde b^\mu$ and $\tilde u^\mu$ together  is given by
\bea
C^{\mn}=\frac{\tilde u^\mu \tilde b^\nu +\tilde b^\mu \tilde u^\nu}{\sqrt{\tilde u^2}\sqrt{\tilde b^2}}.
\eea
To obtain the rest of the tensors, once again we go through the similar steps: at first we define $R^{\mn}=U^{\mn}-B^{\mn}$, which, in this case, can be viewed as the  $A^{\mn}$ and the  $B^{\mn}$ subtracted part of our vacuum basis tensor $V^{\mn}$. Then we obtain the $\tilde a^\mu$ from $a^\mu$ as $\tilde a^\mu= R^{\mn}a_\nu$. This time, the newly constructed $\tilde a^\mu$ becomes orthogonal to $\tilde b^\mu$ as well as $\tilde u^\mu$. Note that, all the four vectors of the set $\tilde u^\mu$, $\tilde b^\mu$ and   $\tilde a^\mu$ are orthogonal to the gluon four momentum $P^\mu$ and hence the basis tensors constructed using them trivially satisfy the transversality condition. Moreover, because of the orthogonality among the constructed  four vectors, the extraction of the form factors gets simplified considerably.  
Now, with the constructed set, the  rest of the independent symmetric  tensors can also  be obtained intuitively as
\bea
D^{\mn}&=& \frac{\tilde a^\mu \tilde a^\nu}{\tilde a^2},\\
E^{\mn}&=& \frac{\tilde u^\mu \tilde a^\nu +\tilde a^\mu \tilde u^\nu}{\sqrt{\tilde u^2}\sqrt{\tilde a^2}},\\
F^{\mn}&=& \frac{\tilde a^\mu \tilde b^\nu +\tilde b^\mu \tilde a^\nu}{\sqrt{\tilde a^2}\sqrt{\tilde b^2}}.
\eea
 %The tensors satisfy following properties
\iffalse
\textcolor{blue}
{
\bea
B^{\mu \rho}Q_{\rho \nu}&=&0,\\
B^{\mu \rho}C_{\rho \nu}&=&0,\\
C^{\mu \rho}Q_{\rho \nu}&=&0,\\
N^{\mu \rho}N_{\rho \nu}&=& Q^\mu_\nu +B^\mu_\nu,\\
E^{\mu \rho}E_{\rho \nu}&=& B^\mu_\nu +C^\mu_\nu,\\
F^{\mu \rho}F_{\rho \nu}&=& C^\mu_\nu +Q^\mu_\nu,\\
B^{\mu \rho}N_{\rho \nu}&+&N^{\mu \rho}B_{\rho \nu}=N^\mu_\nu,\\
Q^{\mu \rho}N_{\rho \nu}&+&N^{\mu \rho}Q_{\rho \nu}=N^\mu_\nu,\\
B^{\mu \rho}E_{\rho \nu}&+&E^{\mu \rho}B_{\rho \nu}=E^\mu_\nu,\\
C^{\mu \rho}E_{\rho \nu}&+&E^{\mu \rho}C_{\rho \nu}=E^\mu_\nu,\\
Q^{\mu \rho}F_{\rho \nu}&+&F^{\mu \rho}Q_{\rho \nu}=F^\mu_\nu,\\
C^{\mu \rho}F_{\rho \nu}&+&F^{\mu \rho}C_{\rho \nu}=F^\mu_\nu,\\
N^{\mu \rho}F_{\rho \nu}&+&F^{\mu \rho}N_{\rho \nu}=E^\mu_\nu,\\
E^{\mu \rho}N_{\rho \nu}&+&N^{\mu \rho}E_{\rho \nu}=F^\mu_\nu,\\
F^{\mu \rho}E_{\rho \nu}&+&E^{\mu \rho}F_{\rho \nu}=N^\mu_\nu,\\
B^\mu_\mu=1\,,\,Q^\mu_\mu=1\,&,&\,C^\mu_\mu=1\,,\,D^\mu_\mu=0,\\
N^\mu_\mu=0\,,\, E^\mu_\mu=0\,&,&\,F^\mu_\mu=0.
\eea
}
\fi
The general structure of the gauge boson self-energy in presence of an ellipsoidal  anisotropic  medium can be expressed as a linear combination of the six basis tensors as
\iffalse
\textcolor{blue}{
\bea
\Pi^{\mn}=\alpha B^{\mn}+\beta Q^{\mn}+\gamma C^{\mn}+\delta D^{\mn}+\rho N^{\mn}+\sigma E^{\mn}+\lambda F^{\mn}.
\eea
}
\fi
\bea
\Pi^{\mn}&=&\alpha A^{\mn}+\beta B^{\mn}+\gamma C^{\mn}+\delta D^{\mn}+\sigma E^{\mn}+\lambda F^{\mn}~.
\eea
It is worth mentioning here that though  tensors like $V^{\mn}$ are used to obtain the set, once we declare our  constructed set of tensors as independent, all the other symmetric and transverse tensors not belonging to the set  become  expressible as their linear combination. For example, it can be shown that  

\bea
V^{\mn}&=& A^{\mn} + B^{\mn} + D^{\mn}.
\eea
Now, we can obtain the  effective propagator from the Dyson-Schwinger equation 

\bea
i\mathcal{D}^{\mn}&=&i\mathcal{D}_0^{\mn}+i\mathcal{D}_0^{\mu\rho} (i\Pi_{\rho\rho'})i\mathcal{D}^{\rho'\nu},
\eea
where the inverse bare propagator $\mathcal{D}_0^{\mn}$ without the explicit color indices is given by 
\bea
(\mathcal{D}^{-1}_0)^{\mn}&=&-P^2\eta^{\mn}-\frac{1-\zeta}{\zeta}P^\mu P^\nu ,
\eea

with $\zeta$ representing the gauge fixing parameter.  
To obtain the effective propagator, let us first consider  the inner product identities   among the basis tensors. To express the identities in a compact form,  here we suppress the Lorentz indices of the basis tensors. Also  the tensors $C$, $E$, and $F$ are considered as a sum of two parts, for example  $C=\overline{C}+\underline{C}$ and in similar fashion for the other two  where 
\bea 
\overline{C}_\nu^\mu=\frac{\tilde u^\mu \tilde b_\nu}{\sqrt{\tilde u^2}\sqrt{\tilde b^2}}, \hspace{0.5cm} \underline{C}_\nu^\mu=\frac{ \tilde b^\mu \tilde u_\nu}{\sqrt{\tilde u^2}\sqrt{\tilde b^2}},\nn\\
\overline{E}_\nu^\mu=\frac{\tilde u^\mu \tilde a_\nu}{\sqrt{\tilde u^2}\sqrt{\tilde a^2}}, \hspace{0.5cm} \underline{E}_\nu^\mu=\frac{ \tilde a^\mu \tilde u_\nu}{\sqrt{\tilde u^2}\sqrt{\tilde a^2}},\nn\\
\overline{F}_\nu^\mu=\frac{\tilde a^\mu \tilde b_\nu}{\sqrt{\tilde a^2}\sqrt{\tilde b^2}}, \hspace{0.5cm} \underline{F}_\nu^\mu=\frac{ \tilde b^\mu \tilde a_\nu}{\sqrt{\tilde a^2}\sqrt{\tilde b^2}}.\nn
\eea
With this notation, the inner product between the basis tensors can be written in a compact form as 
\bea
\renewcommand*{\arraystretch}{1.2}
\centering
 \begin{tabular}{ |c !{\vrule width0.8pt} c c c c c c| }
\hline 
 & $A$ & $B$ & $C$ & $D$ & $E$ & $F$   \\ \thickhline
 $A$~ & $A$ & 0 & $\overline{C}$ & 0 & $\overline{E}$ & 0   \\ 
 $B$~ & 0 & $B$ & $\underline{C}$ & 0 & 0 & $\underline{F}$   \\
 $C$~ & $\underline{C}$ & $\overline{C}$ & $A+B$ & 0 & $\underline{F}$ & $\overline{E}$   \\  
 $D$~ & 0 & 0 & 0 & $D$ & $\underline{E}$ & $\overline{F}$   \\
  $E$~ & $\underline{E}$ & 0 & $\overline{F}$ & $\overline{E}$ & $A+D$ & $\overline{C}$   \\
 $F$~ &0 & $\overline{F}$ & $\underline{E}$ & $\underline{F}$ & $\underline{C}$ & $B+D$   \\\hline
 \end{tabular} 
 \label{contract_1}
\eea
where the composition rule for each entry of the multiplication table  is defined as $r^{\mu}_{\rho} c^{\rho}_{\nu}$ with $r$ and $c$ being members representing  the row and the column respectively. Moreover, further contraction of the free indices as $r^{\mu}_{\rho} c^{\rho}_{\mu}$ simplifies to 

\bea
\renewcommand*{\arraystretch}{1.2}
\centering
 \begin{tabular}{ |c !{\vrule width0.8pt} c c c c c c| }
 \hline 
 &  $A$ & $B$ & $C$ & $D$ & $E$ & $F$   \\ \thickhline
 $A$~ & 1 & 0 & 0 & 0 & 0 & 0   \\ 
 $B$~ & 0 & 1 & 0 & 0 & 0 & 0   \\
 $C$~ & 0 & 0 & 2 & 0 & 0 & 0  \\  
 $D$~ & 0 & 0 & 0 & 1 & 0 & 0   \\
 $E$~ & 0 & 0 & 0 & 0 & 2 & 0  \\
 $F$~ &0 & 0 & 0 & 0 & 0 & 2   \\\hline
  \end{tabular} 
   \label{contract_2}
 \eea
 which will be useful to extract the coefficients of the basis tensors, {\it{i.e.,}} the form factors from the polarization function. Using the properties of the basis tensors from Eq.~\eqref{contract_1} one can obtain the effective gluon propagator given by 
\begin{small}
\bea
\mathcal D^{\mn}&=&-\frac{\beta \delta-\left(\beta +\delta \right) P^2-\lambda^2+P^4}{\Delta}A^{\mn}
   -\frac{ \delta \alpha -\left(\delta + \alpha  \right)P^2-\sigma^2+P^4 }{\Delta}B^{\mn}
   -\frac{\gamma \left(P^2-\delta\right)+\sigma \lambda}{\Delta}C^{\mn}\nn\\
   &-&\frac{\alpha \beta-\left(\alpha +\beta \right) P^2-\gamma^2+P^4 }{\Delta}D^{\mn}
   -\frac{\sigma\left( P^2-\beta\right)+ \lambda \gamma}{\Delta}E^{\mn}
   - \frac{\lambda \left(P^2-\alpha\right)+\gamma \sigma }{\Delta}F^{\mn}
    -\zeta\frac{ P^\mu P^\nu}{P^4}.
    \label{eff_prop}
\eea
\end{small}
where the common denominator of the basis tensors, $\Delta$ is given by 
\bea
\Delta&=&P^6-(\alpha + \beta +  \delta) P^4- ( \gamma^2  +\sigma^2+\lambda^2-\alpha \beta -\beta \delta  -  \delta \alpha ) P^2+\alpha\lambda^2+\beta \sigma^2+\delta\gamma^2 -\alpha \beta\delta-2 \gamma \sigma \lambda.
\eea
Notice that, in presence of ellipsoidal anisotropy, the denominator of the effective propagator becomes a cubic equation of $P^2$ and can be written as a product of three factors as 
\bea
\Delta&=&(P^2-\Omega_0)(P^2-\Omega_+)(P^2-\Omega_-),
\label{disp_modes}
\eea
where $\Omega_0$ and  $\Omega_\pm$ can be written in terms of the form factors as 
\bea
\Omega_0&=&\frac{1}{3}(\alpha + \beta + \delta)- \frac{1}{3}\frac{\vp}{\Big(\frac{\chi +\sqrt{4\vp^3+\chi^2}}{2}\Big)^{\frac{1}{3}}}+\frac{1}{3}\Big(\frac{\chi +\sqrt{4\vp^3+\chi^2}}{2}\Big)^{\frac{1}{3}},\label{om0}\\
\Omega_+&=&\frac{1}{3}(\alpha + \beta + \delta)+ \frac{1+i\sqrt{3}}{6}\frac{\vp}{\Big(\frac{\chi +\sqrt{4\vp^3+\chi^2}}{2}\Big)^{\frac{1}{3}}}-\frac{1-i\sqrt{3}}{6}\Big(\frac{\chi +\sqrt{4\vp^3+\chi^2}}{2}\Big)^{\frac{1}{3}},\label{omp}\\
\Omega_-&=&\frac{1}{3}(\alpha + \beta + \delta)+ \frac{1-i\sqrt{3}}{6}\frac{\vp}{\Big(\frac{\chi +\sqrt{4\vp^3+\chi^2}}{2}\Big)^{\frac{1}{3}}}-\frac{1+i\sqrt{3}}{6}\Big(\frac{\chi +\sqrt{4\vp^3+\chi^2}}{2}\Big)^{\frac{1}{3}},\label{omm}
\eea
where $\vp$ and $\chi$ are given by 
\bea
\vp&=&\alpha(\beta-\alpha)+\beta(\delta-\beta)+\delta(\alpha-\delta)-3(\gamma^2+\lambda^2+\sigma^2),\\
\chi&=&(2\alpha-\beta-\delta)(2\beta-\delta-\alpha)(2\delta-\alpha-\beta)+54\gamma\lambda\sigma\nn\\
&-&9\alpha(2\lambda^2-\sigma^2-\gamma^2)-9\beta(2\sigma^2-\gamma^2-\lambda^2)
-9\delta(2\gamma^2-\lambda^2-\sigma^2).
\eea
  Once the form factors are extracted from the polarization function, the desired collective modes of the gluon can be obtained from the pole of the effective propagator.

\section{Results}\label{sec.results}
\begin{figure}
\begin{center}
 \includegraphics[scale=0.45]{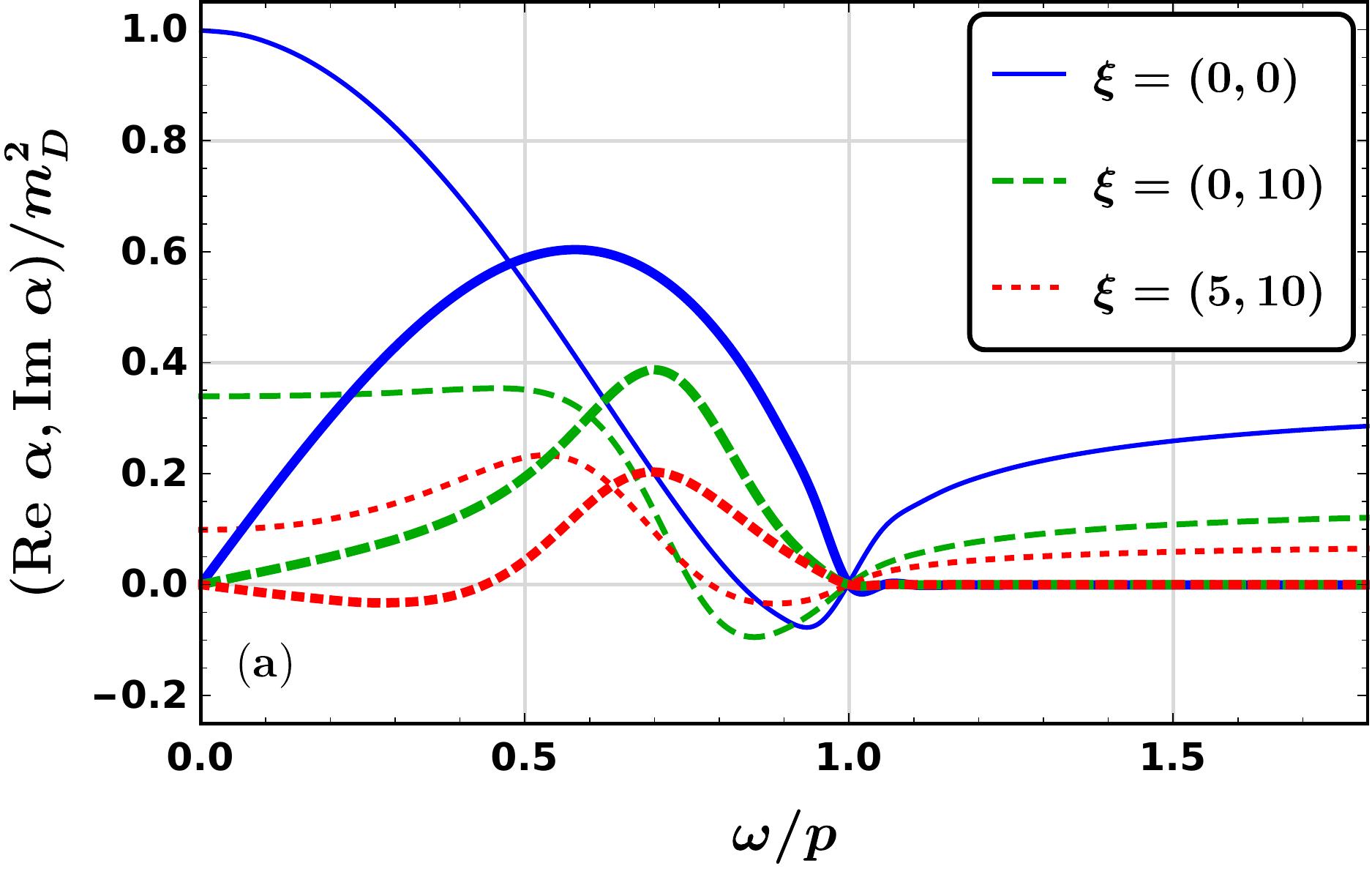}
 \includegraphics[scale=0.45]{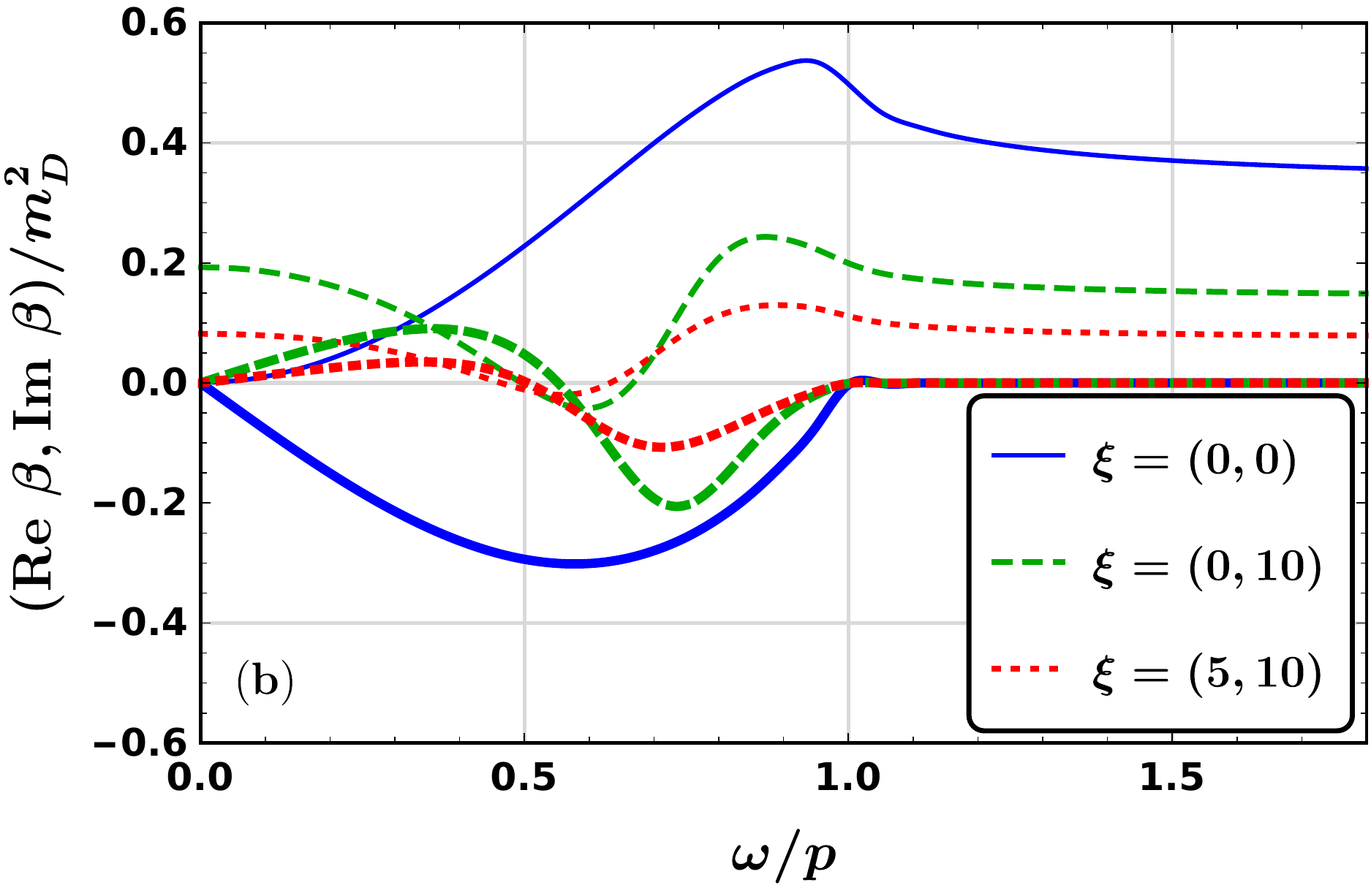}
 \includegraphics[scale=0.45]{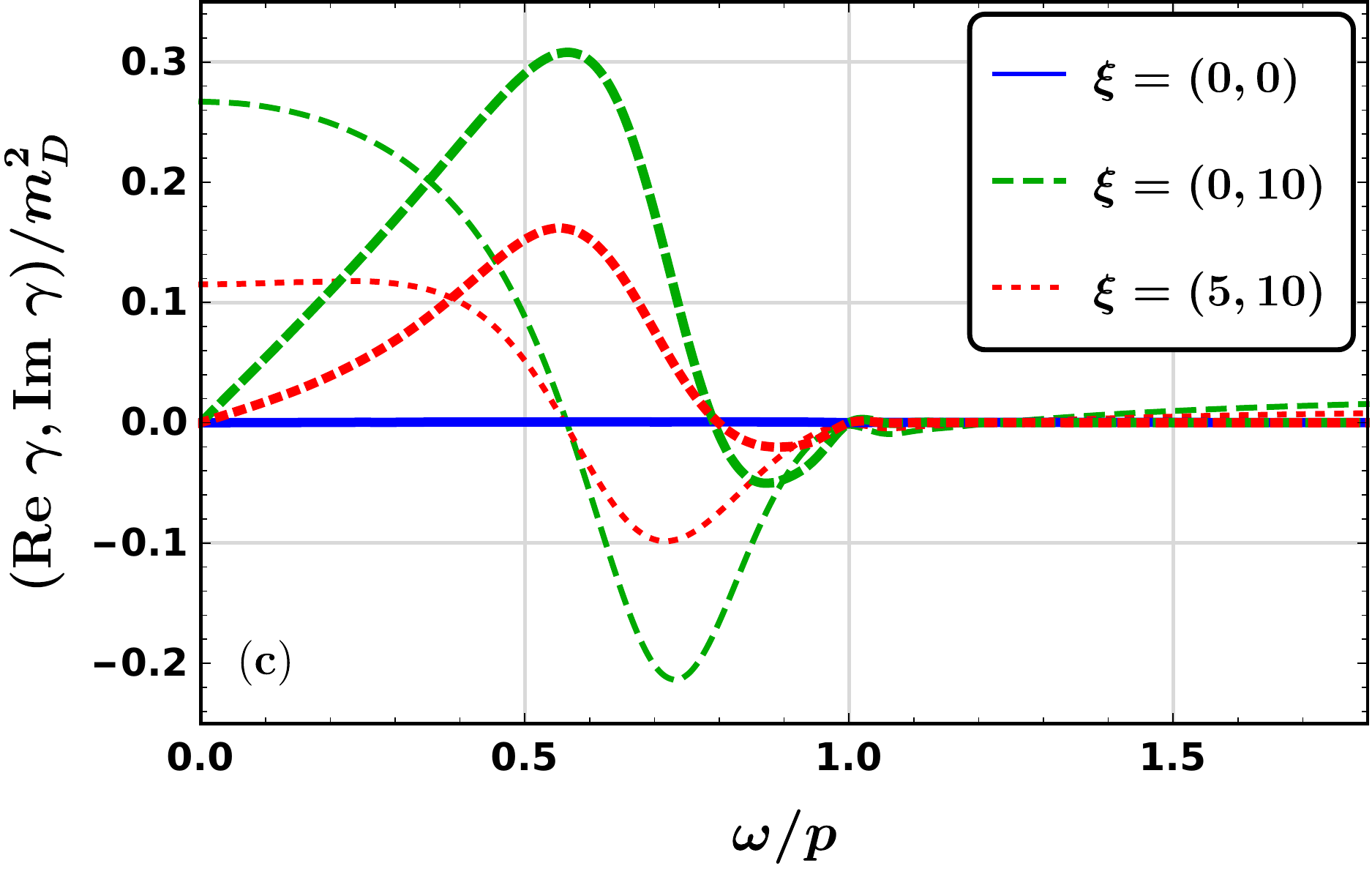}
 \includegraphics[scale=0.45]{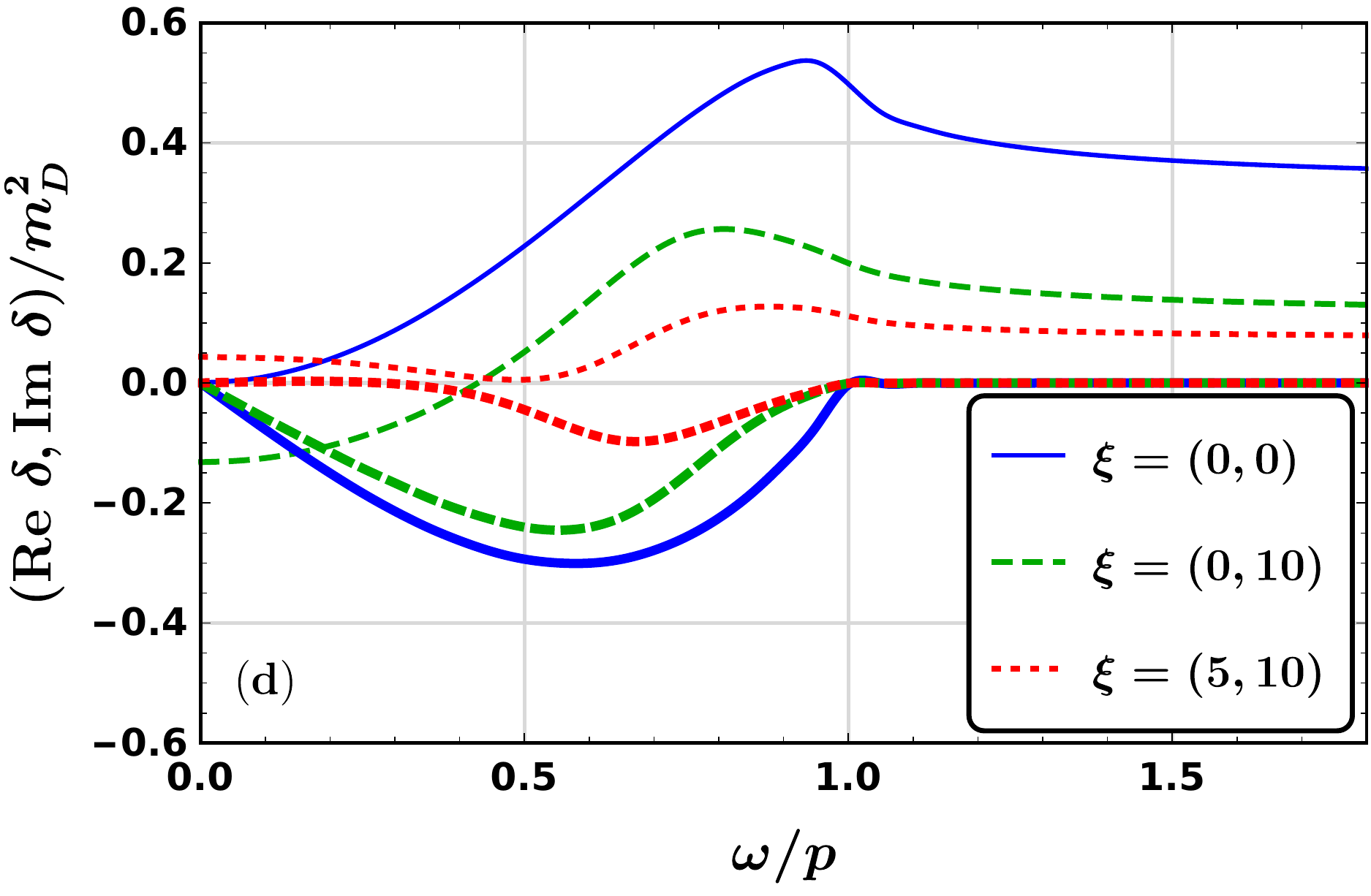}
 \includegraphics[scale=0.45]{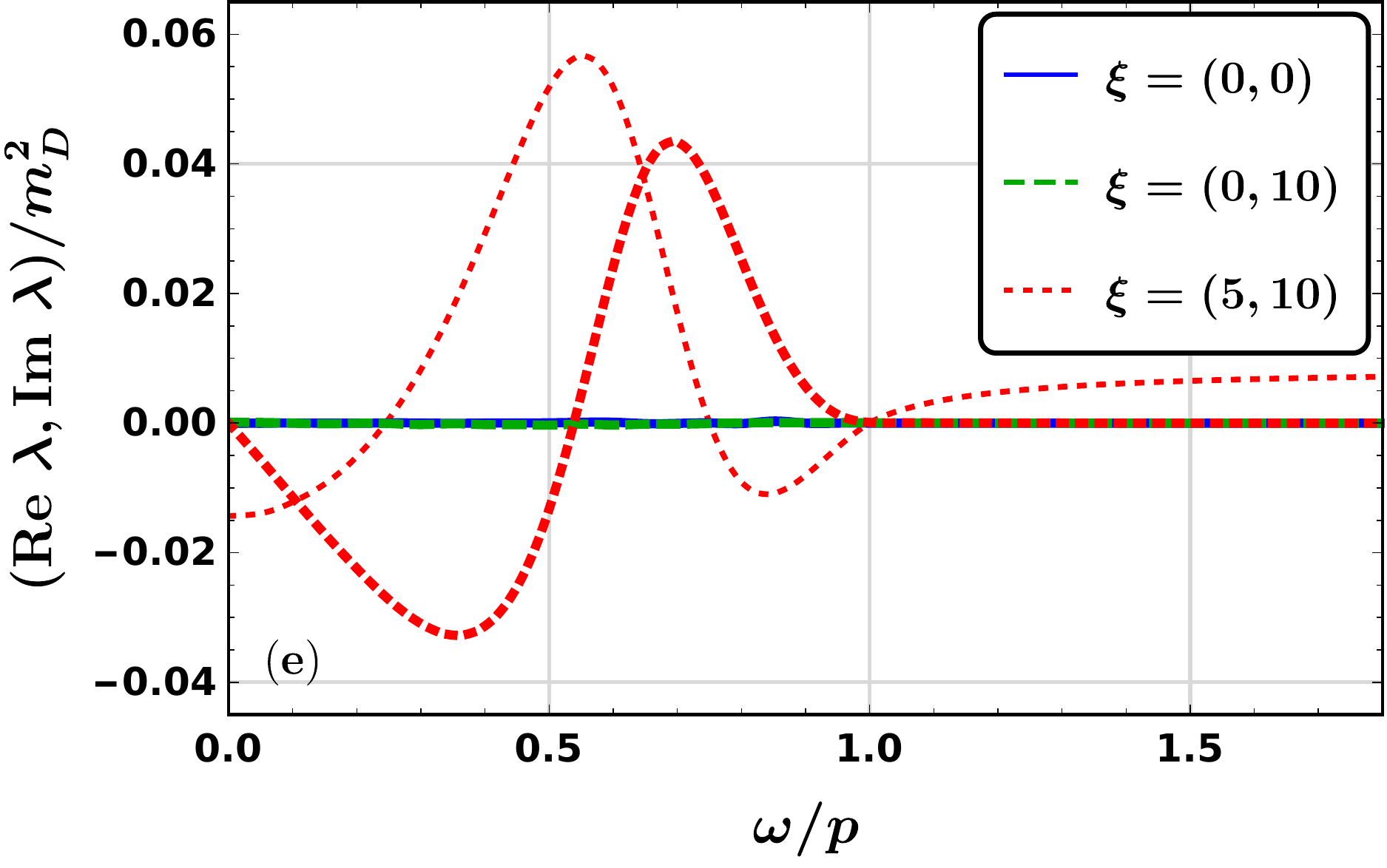}
 \includegraphics[scale=0.45]{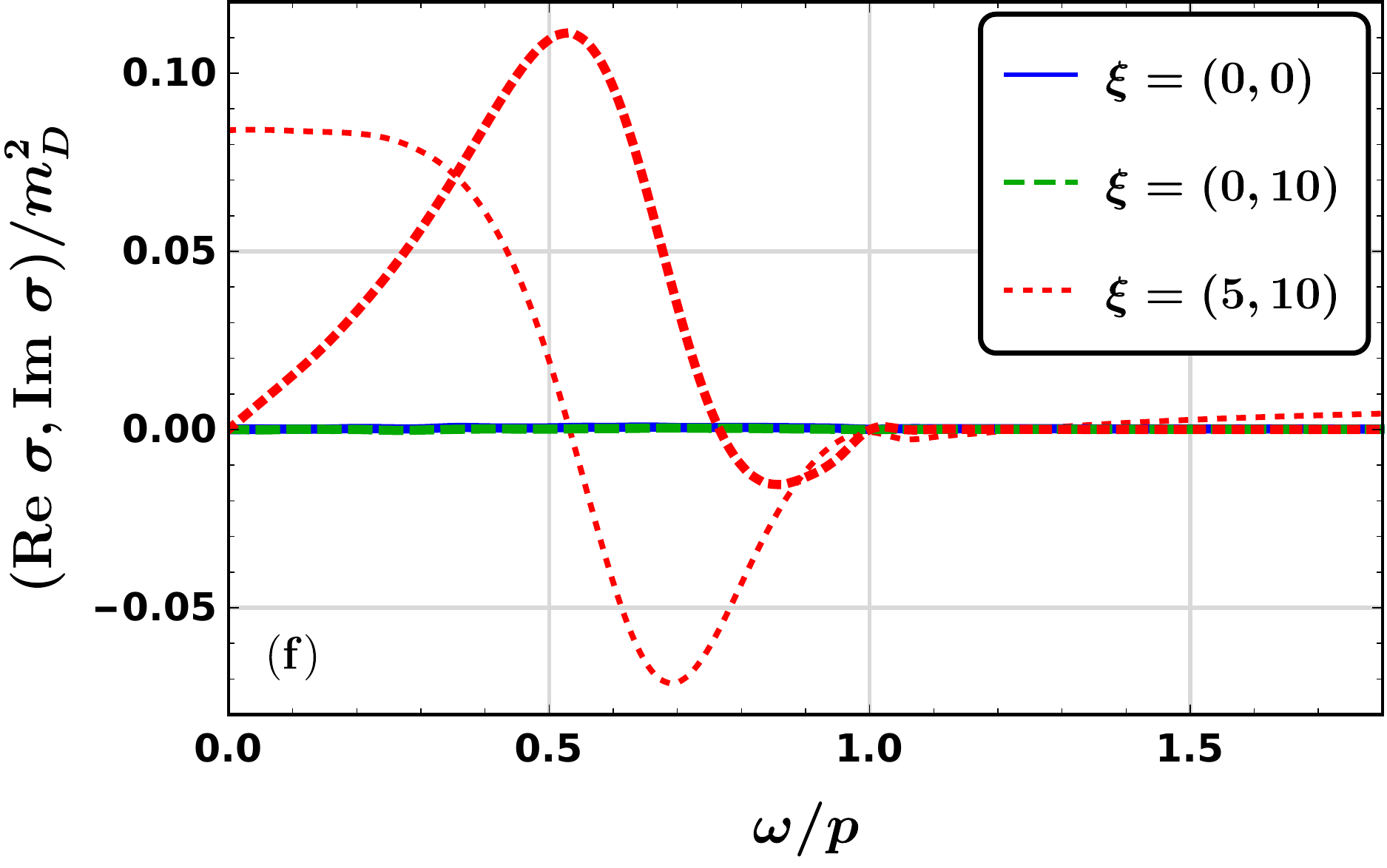}
 \caption{The real and imaginary parts of the form factors are plotted in (a)--(f) as a function of $\omega/p$ for three different sets of anistotropy tuple $\xi=(\xi_a,\xi_b)=(0,0),(0,10),$ and $(5,10)$. In each case, the imaginary parts are shown with comparatively  thicker style. The spheroidal set is shown  at fixed $\theta_p=\pi/4$ whereas for the ellipsoidal case, $(\theta_p,\phi_p)=(\pi/4,\pi/3)$ is considered.  }
  \label{formfactors}
\end{center} 
\end{figure}

In this work we consider the hard-loop gluon polarization tensor in a nonequilibrium QGP medium given by \cite{Mrowczynski:2000ed}

\bea
\Pi^{\mu\nu}(p^0=\omega, {\bm p})&=&g_s^2\int\frac{d^3{\bm k}}{(2\pi)^3}\frac{K^\mu}{E_k}\frac{\partial f({ \bm k})}{\partial K^\rho}\Big[\eta^{\rho\nu}-\frac{K^\rho P^\nu}{P\cdot K+i0^+}\Big],
\eea
where $g_s$ is the strong coupling constant and $k^0=E_k$ represents the energy of the massless partonic degrees of freedom that  modify the gluon dispersion in presence of anisotropic medium.  The effective distribution function $f({ \bm k})$ is given by 
\bea
f({ \bm k})=2N_c n_g({\bm k})+N_f\big[n_q({\bm k})+n_{\overline{q}}({\bm k})\big],
\eea
where $n_g({\bm k})$ represents the gluon number density whereas $n_q({\bm k})$ and $n_{\overline{q}}({\bm k})$ are the quark and antiquark number densities respectively. $N_c$ and $N_f$ are respectively the number of colors and quark flavors.  
A general method of constructing anisotropic momentum distribution function is to transform the argument of an isotropic momentum space distribution essentially by introducing  deformations in a parametrized way. A discussion on the generalization of the RS form to ellipsoidal anisotropies can be found in Refs.\cite{Nopoush:2014pfa,Kasmaei:2018yrr}.   In our case we use the  ellipsoidal momentum distribution parametrized as \cite{Kasmaei:2016apv,Kasmaei:2018yrr}
\bea
f_{\mbox{aniso}}({\bm k})\equiv f_{\mbox{iso}}\big(\frac{1}{\Lambda}\sqrt{{\bm k}^2+\xi_a({\bm k}\cdot{\bm a})^2+\xi_b({\bm k}\cdot{\bm b})^2}\big),
\eea
where $\Lambda$ represents a temperature-like scale which, in the equilibrium limit, corresponds to the temperature. The  gluon polarization tensor with such anisotropic parton distributions  can be written as
\bea
\Pi^{\mu\nu}(\omega,{\bm p},\xi)&=&m_D^2\int\frac{d\Omega}{4\pi}v^\mu\frac{v^l+\xi_a({\bm v}\cdot {\bm a})a^l+\xi_b({\bm v}\cdot {\bm b})b^l}{(1+\xi_a({\bm v}\cdot {\bm a})^2+\xi_b({\bm v}\cdot {\bm b})^2)^2}\Big[\eta^{\nu l}-\frac{v^\nu P^l}{\omega-{\bm p}\cdot {\bm v}+i0^+}\Big],\label{polarization tensor}
\eea
where $\xi$ represents the anisotropy tuple $(\xi_a,\xi_b)$ and   $m_D^2=(N_c+N_f/2)\frac{g_s^2\Lambda^2}{3}$ corresponds to the QCD Debye mass scale. In the above expression, the parton four velocity $v^\mu=K^\mu /k$   and the components of    ${\bm a}$, ${\bm b}$, ${\bm v}$, and ${\bm p}$  in the rest frame of the medium are chosen  as 
\bea
{\bm a}&=&(1,0,0),\\
{\bm b}&=&(0,0,1),\\
{\bm v}&=&(\sin\theta_k \cos\phi_k,\sin\theta_k \sin\phi_k,\cos\theta_k),\\
{\bm p}&=&p(\sin\theta_p \cos\phi_p,\sin\theta_p \sin\phi_p,\cos\theta_p),\label{coordinate}
\eea
with $d\Omega$ representing the differential solid angle corresponding to the internal angular coordinates $(\theta_k, \phi_k)$. It should be mentioned here  that another common choice of reference frame is the  parton specific coordinate as used in Ref.~\cite{Kasmaei:2018yrr} where one reorients the axes so that the polar angle is measured  with respect to the parton momentum. However, as we are interested in the evaluation of the  form factors, the results do not depend on any  specific choice of the reference frame.

Now, as discussed in the previous section, the gluon self-energy  can be decomposed in terms of six independent basis tensors.  Utilizing the contraction relations among the basis tensors, the corresponding form factors can be obtained in terms of the self-energy components  given in Eq.\eqref{polarization tensor} as a two dimensional integral over the solid angle. However, because of the  transversality condition and the symmetry under the exchange of the free indices, the number of independent components of the self-energy reduces to six. Thus, any chosen set of six independent components  will be sufficient to determine the form factors. It should be mentioned here that, though in our work, the form factors are obtained  numerically using the conventional quadrature routines, an alternative way involving the hypergeometric  expansion method is expected to be much more efficient \cite{Kasmaei:2018yrr} for this purpose.   
The  real and imaginary parts of the form factors are shown in Fig.~\ref{formfactors}, for   three different situations, namely the isotropic case with  $\xi=(0,0)$, the spheroidal case with $\xi=(0,10)$ and the ellipsoidal case with  $\xi=(5,10)$. The spheroidal case is shown for $\theta_p=\pi/4$ whereas for the ellipsoidal anisotropy,  $\theta_p$ and $\phi_p$, are chosen as $\pi/4$ and $\pi/3$, respectively. 
 The imaginary parts of the form factors in all  the three cases  exist only in the spacelike region. 
 It should be noted that, when the isotropic medium is considered,  $\beta$ and $\delta$ become degenerate whereas $\gamma$, $\lambda$, and $\sigma$ become zero. This results in two distinct dispersive modes of the gluon among which one is degenerate, {\it{i.e.,}} $\Omega_0=\alpha$ and $\Omega_+=\Omega_-=\beta=\delta$.
The analytic expression of the degenerate form factors is given by
\bea
\beta=\delta=\Pi_T=\frac{m_D^2}{2}\frac{\omega^2}{p^2}\bigg[1-\frac{\omega^2-p^2}{2\omega p}\ln \frac{\omega+p}{\omega-p} \bigg],
\eea
whereas the dispersion for the other distinct mode can be obtained from
\bea
\alpha=\Pi_L=\frac{m_D^2}{ \tilde{u}^2}\bigg[ 1-\frac{\omega}{2p}\ln \frac{\omega+p}{\omega-p}\bigg],
\eea
which are the familiar results of the gluon self-energy in isotropic thermal medium~\cite{Bellac:2011kqa}. Here $m_D^2=(N_c+N_f/2)\frac{g_s^2 T^2}{3}$ and $\tilde{u}^2=-p^2/P^2$.  In the presence of spheroidal anisotropy, $\sigma$ and $\lambda$ remain zero. However, $\beta$ and $\delta$ are no longer degenerate. In that case the functions  corresponding to the dispersive modes   simplify to
 \bea
 \Omega_0&=&  \frac{1}{2}\bigg( \alpha + \beta+\sqrt{(\alpha - \beta)^2+4\gamma^2} \bigg), \label{RS1}\\
  \Omega_+&=& \frac{1}{2}\bigg( \alpha + \beta-\sqrt{(\alpha - \beta)^2+4\gamma^2} \bigg), \label{RS2}\\
   \Omega_-&=& \delta \label{RS3}.
 \eea
which may be compared with the modes obtained  in Ref.~\cite{Romatschke:2003ms} 
(see for example Eq.(43) therein). Though we find a different combination of the form factors in the  expressions due to the different choice of our basis tensors, it should be noticed that, in our case too, the arguments inside the square root appear as a sum of two complete squares, thereby  allowing similar interpretation for the  collective modes [see for example Sec. VI of \cite{Romatschke:2003ms}]. 
  When ellipsoidal anisotropy is considered, it is observed that  all the form factors are nonzero and they contribute in the gluon dispersion. However, It should be noticed that, for the fixed set of parameters  chosen for the figure, the values of $\lambda$ and $\sigma$ are an order of magnitude smaller than the other form factors.

Now, following Ref.~\cite{Romatschke:2003ms}, a  mass scale corresponding to a given form factor, say for example  $\alpha$, can be defined in the static limit as
\bea
m_\alpha^2=\lim_{\omega\rightarrow0} \alpha.
\eea
It is evident from Fig.~\ref{formfactors} that the imaginary part of each of the form factors vanishes at $\omega \rightarrow 0$ limit. Thus, the  mass scales as defined above are real quantities. Now, in a similar way, one can  define the  mass scales corresponding to the gluon dispersive modes as
\bea
m_{\Omega_{0,\pm}}^2=\lim_{\omega\rightarrow0} \Omega_{0,\pm}(\omega,p,\theta_p,\phi_p).
\eea

\begin{figure}
\begin{center}
 \includegraphics[height=6.15cm,width=8cm]{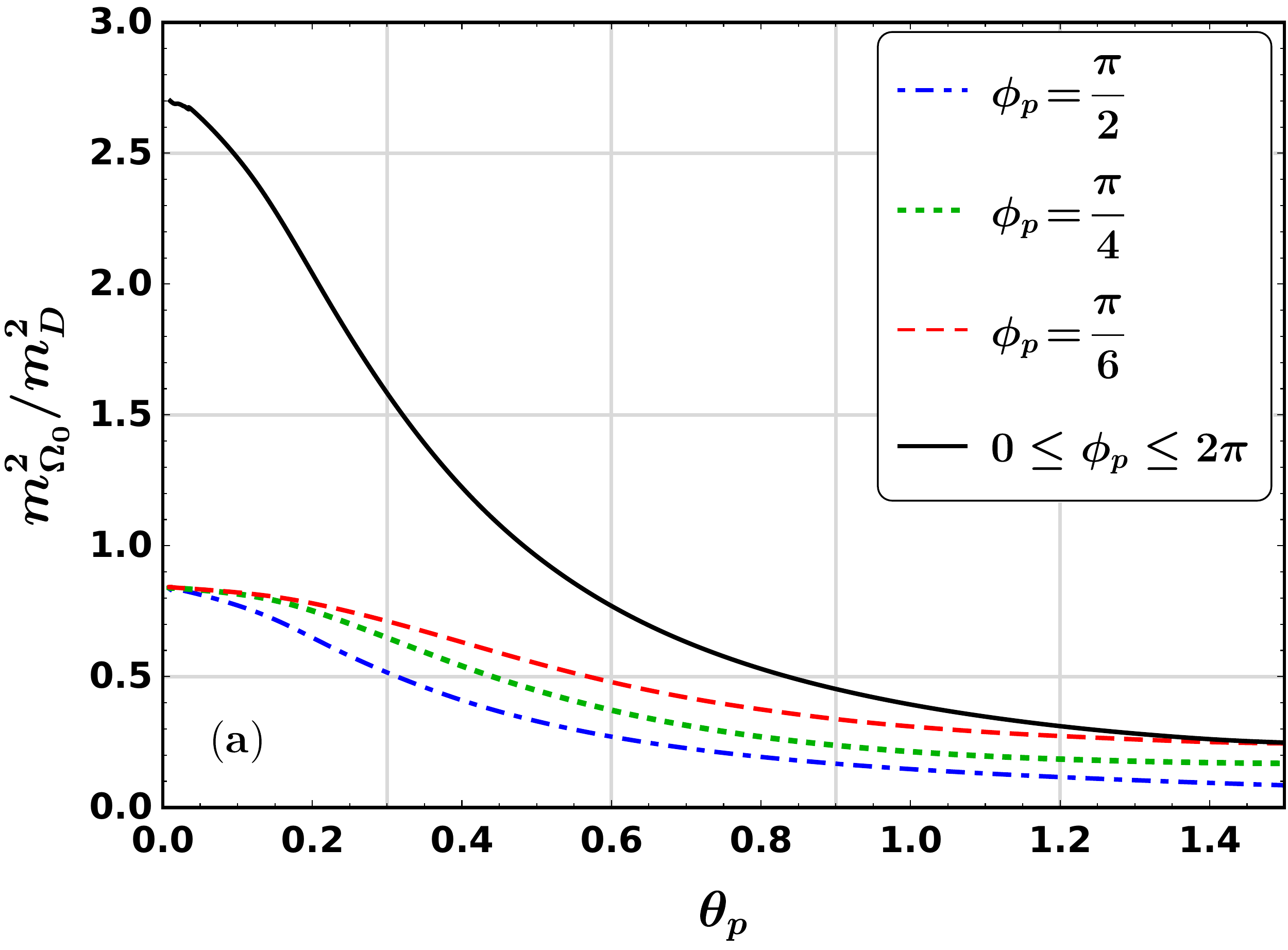}
 \includegraphics[height=6cm,width=8cm]{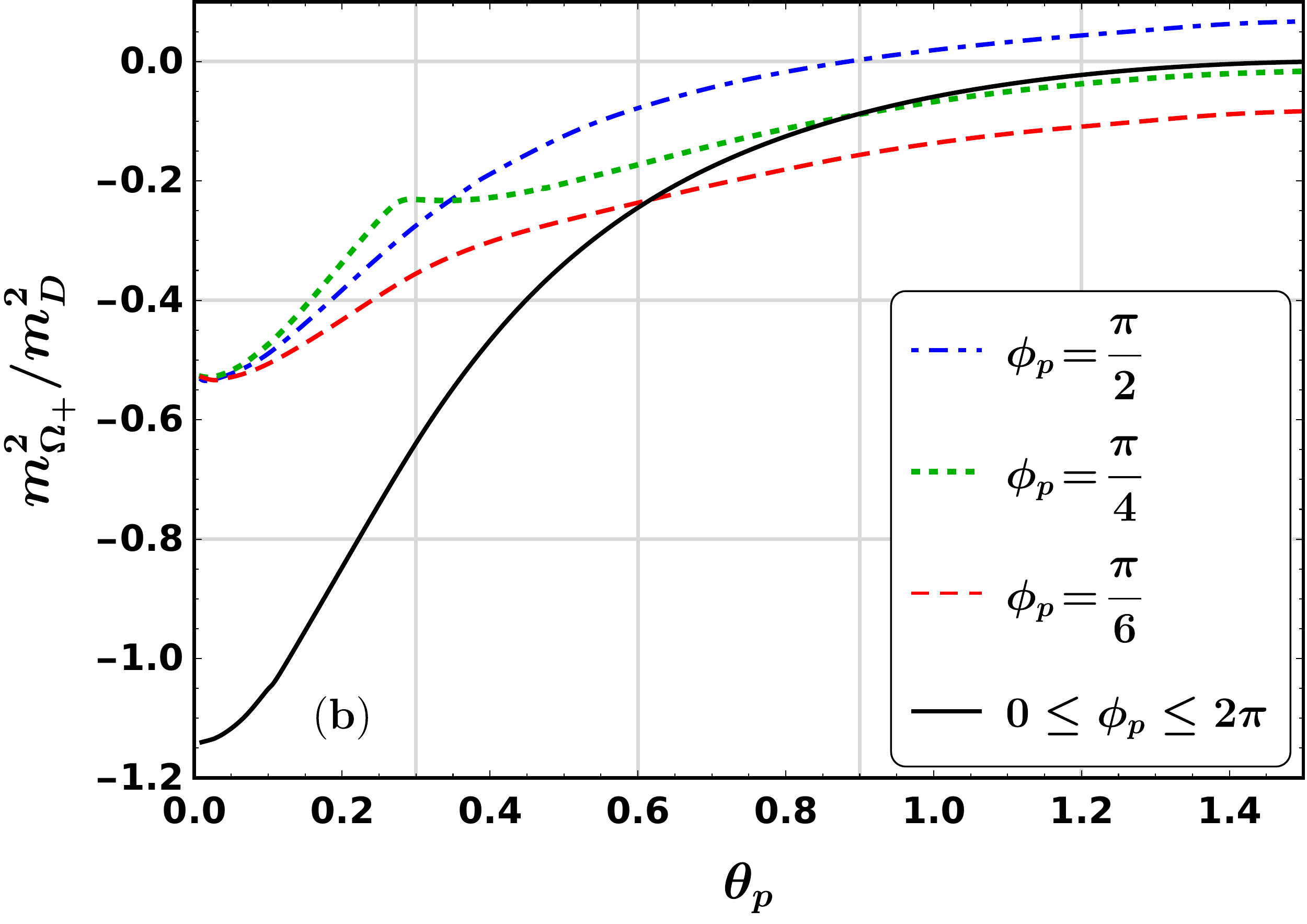}
 \includegraphics[height=6cm,width=8cm]{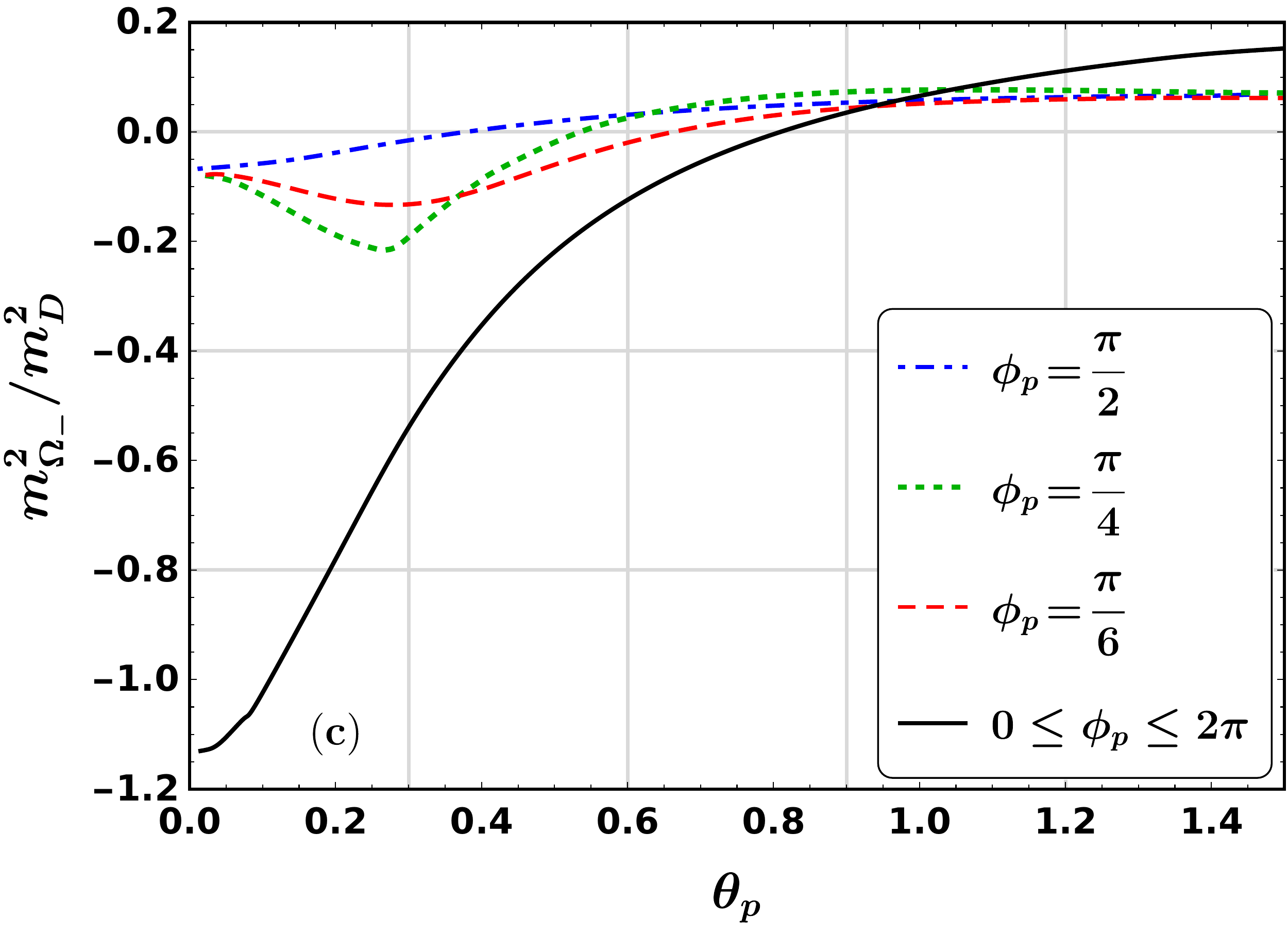}
    \caption{Squared values of the mass scales corresponding to  $\Omega_0$ and $\Omega_\pm$ are plotted in (a)--(c) as a function of  $\theta_p$ for three different values of $\phi_p= \{\pi/2,\pi/4,\pi/6\}$ (shown in discrete style) considering $(\xi_a,\xi_b)=(5,10)$. The solid line represents the spheroidal case with $(\xi_a,\xi_b)=(0,10)$.  }
  \label{mass_plot}
\end{center} 
\end{figure}

\begin{figure}
\begin{center}  
  \includegraphics[height=6cm,width=8cm]{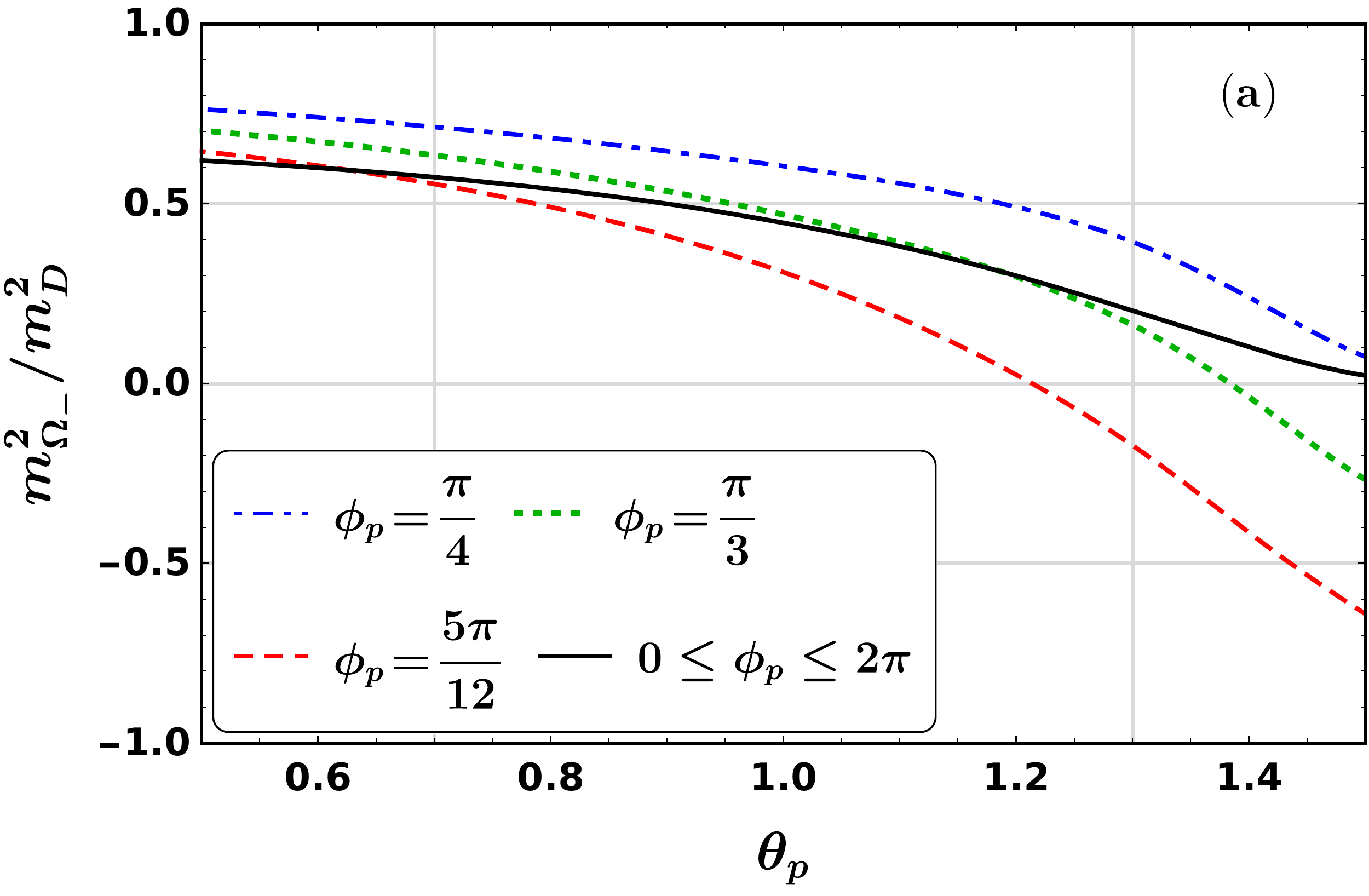}  
  \includegraphics[height=6cm,width=8cm]{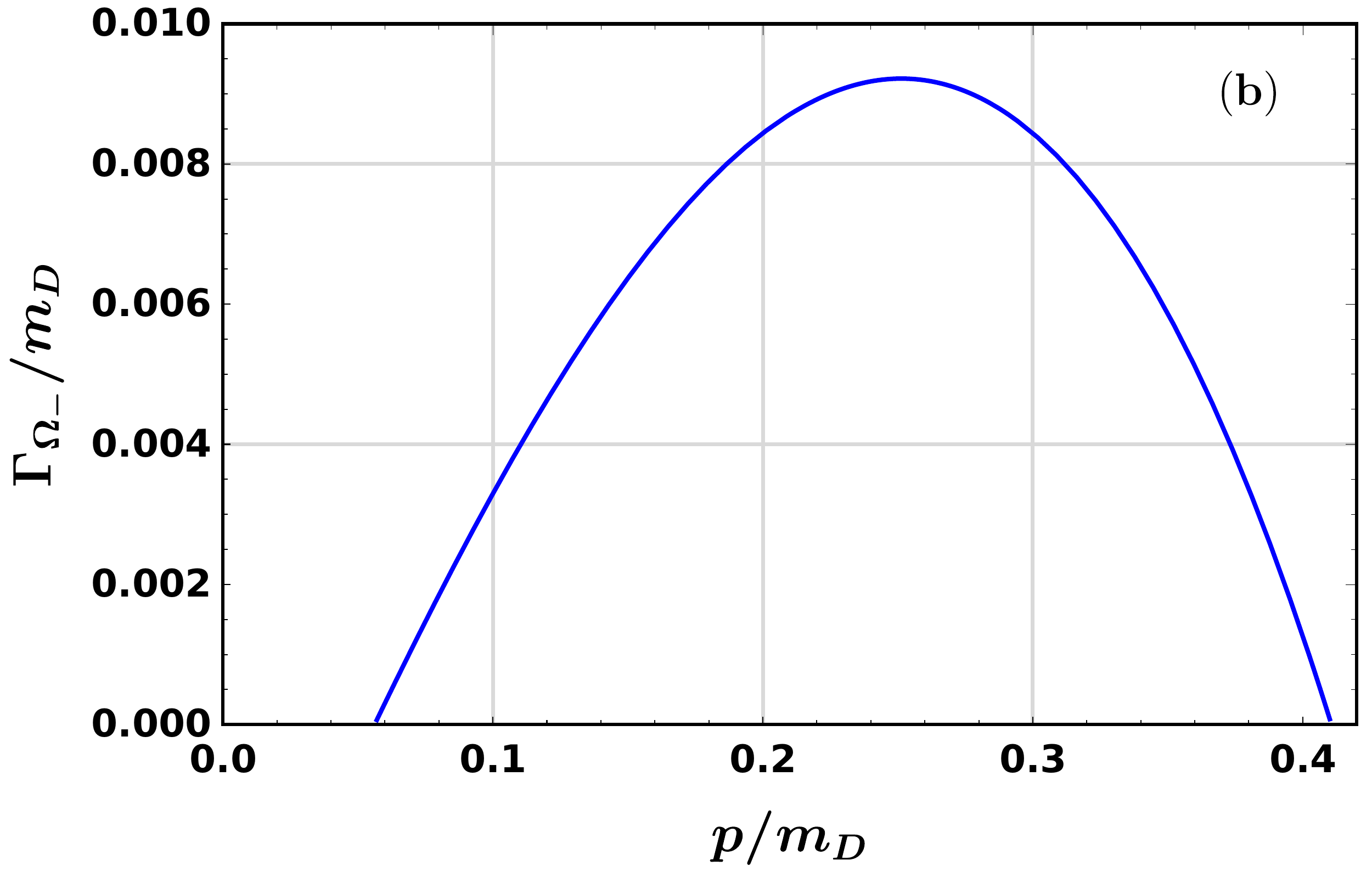}  
   \caption{In the left panel, the squared value of the mass scale corresponding to  $\Omega_-$ is plotted as a function of  $\theta_p$ for three different values of $\phi_p= \{\pi/4,\pi/3,5\pi/12\}$ (shown in discrete style) considering $(\xi_a,\xi_b)=(-0.5,-0.9)$. The  spheroidal case (shown in solid style) with $(\xi_a,\xi_b)=(0,-0.9)$ is also plotted for comparison. In the right panel, the growth rate for the $\Omega_-$ mode  is plotted as a function of  $p/m_D$ with $(\xi_a,\xi_b)=(-0.5,-0.9)$ and $(\theta_p,\phi_p)=(5\pi/12,5\pi/12)$.   }
  \label{inst}
  \end{center}
  \end{figure}

Their variation with the polar angle $\theta_p$   is  shown  in Fig.~\ref{mass_plot} for  three fixed values of $\phi_p=\{\pi/2,\pi/4,\pi/6\}$. For each modes, the corresponding spheroidal version is also shown for comparison. It is evident from the figures that the azimuthal symmetry of the spheroidal case is now broken with the introduction of additional anisotropy direction. Consequently,  a nontrivial $\phi_p$ dependence can be observed  in  the mass scales.    
As can be seen from the figure,  the  mass scale corresponding to $\Omega_0$  remains positive throughout the  range of $\theta_p$ values as also found in  the spheroidal case. Again, similar to the spheroidal anisotropy, negative values in the mass scale is observed  for $\Omega_\pm$ which correspond to instability ~\cite{Romatschke:2003ms}. However, it can be noticed that as the value of $\phi_p$  approaches to $\pi/2$, $m^2_{\Omega_\pm}$ becomes positive at smaller values of $\theta_p$. In other words, with larger deviation from  azimuthal anisotropy direction, the collective modes can be unstable only in shorter window of $\theta_p$ values as compared to the spheroidal case. However, it should be noted that, the above observation is made with a particular set of anisotropy parameter where both $\xi_a$ and $\xi_b$ are positive. An interesting situation occurs for negative value of anisotropy parameter as shown in the left panel of Fig.~\ref{inst}. Here, the angular variation of  $m_{\Omega_-}^2 $ is shown  with  $(\xi_a,\xi_b)=(-0.5,-0.9)$. In this case, it is observed that, unlike the spheroidal case (which remains stable throughout the $\theta_p$ range), the mass scale corresponding to $\Omega_-$ can be negative, indicating an unstable mode. The corresponding growth rate $\Gamma_{\Omega_-}$ can be obtained by solving the dispersion with the replacement $\omega \rightarrow i \Gamma_{\Omega_-}$, {\it{i.e.,} } by solving the equation 
\bea
(\omega=i\Gamma_{\Omega_-})^2-p^2-\Omega_-(\omega=i \Gamma_{\Omega_-},p,\theta_p,\phi_p)=0. 
\eea
The  corresponding  solution of the growth rate is shown in the right panel of Fig.~\ref{inst} with $\theta_p$ and $\phi_p$ both  fixed at  $5\pi/12$. It should be mentioned here that in this case the growth rate of the  unstable mode has amplitudes similar to the spheroidal case~\cite{Romatschke:2003ms} whereas, with positive anisotropy parameters, a several times larger growth rate can be observed  for certain angular values as reported in Ref.~\cite{Kasmaei:2018yrr}.

\section{Summary}\label{sec.summary}

In this article, we have  studied the  gluon polarization  in presence of  ellipsoidal momentum-space anisotropy.  The momentum distribution function in our case is parametrized with two anisotropy parameters $\xi_a$ and $\xi_b$, represented together as an anisotropy tuple $\xi=(\xi_a,\xi_b)$. This is a simple  generalization of  the spheroidal RS form that has been extensively used  in the literature. The general structure of the gluon polarization tensor in presence of such ellipsoidal anisotropy has been formulated and subsequently,  the gluon effective propagator is obtained. As shown earlier, from the  pole of the effective  propagator, the three collective modes can be obtained in terms of the form factors. The results obtained using our formulation are  in agreement with the previous study incorporating ellipsoidal anisotropy \cite{Kasmaei:2018yrr}. It should be mentioned here that it is not mandatory to consider the general structure of the polarization function  to obtain the collective modes, as, those can also be obtained solving the characteristic equation directly (see for example \cite{Kasmaei:2018yrr}).  However,  
one of the  important advantages of considering the general structure  is that, the collective modes, as in our  case, can be expressed in terms of the  form factors which do not depend on the choice of the frame of reference. As a consequence, it is possible to define mass scales by taking the static limits of the functions  characterizing the collective modes ($\Omega_0$ and $\Omega_\pm$) in a  similar fashion as done in case of spheroidal anisotropy. The importance of such definition lies in the fact that, the existence of instability can be inferred systematically  by studying the angular variations of the mass scales. More specifically, for the given  external angles, the negative value of the squared mass indicates that the corresponding mode is unstable. In our analysis with $\xi_a=-0.5$ and $\xi_b=-0.9$, we have observed  that, unlike the spheroidal case,  the mode corresponding to $\Omega_-$   becomes unstable. The appearance of such additional unstable mode in presence of ellipsoidal anisotropy may have important influences on the  isotropization  of the QGP medium produced in HIC experiments.  As mentioned earlier, the formulation, as developed in this work, will be  particularly useful in the studies concerning the heavy quark potential in presence of  ellipsoidal anisotropy. The usual procedure to obtain the hard-thermal loop resummed perturbative part of the heavy quark potential is to consider the Fourier transform of the  00 component of the effective propagator [obtained in Eq.\eqref{eff_prop}] in the static limits. Consequently, the  nontrivial  angular dependence enters  in the  potential  through the mass scales. It should be noted that, in this work, only the retarded part of the gluon self energy is considered.  On the other hand, the imaginary part of the potential can be obtained from the Feynman effective propagator in the real time Keldysh formalism   \cite{Nopoush:2017zbu}. Due to the azimuthal angular dependence of the mass scales, a nontrivial modification in  the real as well as in the imaginary parts of the heavy quark potential is expected in presence of ellipsoidal anisotropy, which will be  an interesting future direction. 

\section{ACKNOWLEDGMENTS}
R. G. is funded by University Grants Commission
(UGC). B. K. is funded by Department of Atomic
Energy (DAE), India via the project TPAES. A. M.
acknowledges Najmul Haque for immense encouragement
and support. A. M. would like to acknowledge Science and
Engineering Research Board (SERB) for funding.

\end{document}